# Ultrafast modulation in a THz graphene-based flat absorber through negative photoconductivity

**Anna C. Tasolamprou[1], Anastasios D. Koulouklidis[1], Christina Daskalaki[1], Charalampos P. Mavidis[1,2], George Kenanakis[1], George Deligeorgis[1], Zacharias Viskadourakis[1], Polina Kuzhir[3,4], Stelios Tzortzakis[1,2,5], Eleftherios N. Economou[1,6], Maria Kafesaki[1,2] and Costas M. Soukoulis[1,7]**

1. Institute of Electronic Structure and Laser, FORTH, 70013 Heraklion, Crete, Greece
2. Department of Materials Science and Technology, University of Crete, 70013, Heraklion, Crete, Greece
3. Institute for Nuclear Problems, Belarusian State University, Bobruiskaya 11, 220030 Minsk, Belarus
4. Tomsk State University, 36 Lenin Ave., Tomsk 634050, Russia
5. Science Program, Texas A&M University at Qatar, P.O. Box 23874 Doha, Qatar
6. Department of Physics, University of Crete, 70013, Heraklion, Crete, Greece
7. Ames Laboratory and Department of Physics and Astronomy, Iowa State University, Ames, Iowa 50011, United States
*corresponding author, E-mail: atasolam@iesl.forth.gr

## Abstract

We present the experimental and theoretical study of an ultrafast graphene-based thin film absorption modulator for operation in the THz regime. The flat modulator is composed of a graphene-sheet placed on a dielectric layer backed by a metallic back-reflector. A near-IR pulse induces the generation of hot carriers in the graphene sheet reducing effectively its conductivity. The system provides a platform with ultrafast modulation capability for flat optics and graphene-based metasurfaces applications.

## 1. Introduction

Graphene is a two-dimensional material made of carbon atoms arranged in a honeycomb lattice with unique mechanical, thermal, electrical and optical properties. Particularly in the THz spectrum, graphene exhibits a Drude-like response due to its easily generated and controlled free carriers; therefore, it is considered a suitable platform for dynamically tunable metasurface components [1]. Such an example, are the so called thin film absorbers that are structures capable of absorbing all power of incident electromagnetic waves [2]. Within the concept of metasurfaces, thin film absorbers, are usually implemented by placing a lossy material, uniform or with features, on the top of a perfectly conducting metallic plate A simple but yet remarkable graphene component/metasurface is based on the coherent absorption principle. It consists of a single uniform sheet of graphene placed on top of a dielectric film which is placed on a metallic plate [3]. Here we demonstrate experimentally ultrafast (of the order of few ps) THz amplitude modulation of such a graphene based ultra-thin absorber induced by photoexcitation via an optical pump signal [4]. For the experimental characterization we use a broadband THz time-domain-spectroscopic system (THz-TDS) in an IR pump-THz probe configuration. Absorption modulation at 2.17 THz in the order of 40% is reported through a decrease in the conductivity upon photoexitation. Detailed analysis of the experimental observations reveals the negative dynamics of the THz photo-conductivity in the graphene sheet which is observed in highly doped samples [5]. Our study unveils the capability of such an ultrathin system to provide ultrafast modulation appropriate for the demanding future flat optics modulation applications and graphene-based metasurfaces.

## 2. Discussion

In Figure 1(a) we present the schematic of the component under investigation. It consits of an uniform graphene sheet placed on a backplated dielectric substrate. The stuctre is designed to provide enhanced absorption though the conherent intereference of the impinging and backreflected waves at the lossy graphene sheet. This is achieved when the equivlent metasurface is impedance matched with the free space. Impedance match is forced by properly enginnering the detail condition of the graphene and the thickness of the substrate. The modulation of the THz absorption spectrum upon photoexcitation is presented in Figure 1(a). For the characterization we use a powerful THz-TDS system that provides the ability of measurements in reflection mode. It is based on a pump-probe, coherent detection approach, and uses an amplified kHz Ti:Sapphire laser system delivering 35 fs pulses at 800 nm central wavelength and maximum energy of 2.3 mJ/pulse. The initial beam is focused in ambient air after partial frequency doubling in a beta-barium-borate (BBO) crystal (50 m thick) to produce a two-color filament and subsequently, THz radiation (>200 kV/cm). The inset in Figure 1(a) presents the optical pump induced THz relative reflectivity change (DR/R) as a function of pump delay. The measurement refers to the peak of the first pulse reflected by the graphene sheet, prior to cavity. The observed change in the reflectivity points to a reduction of the graphene conductivity in the regime under investigation. This counterintuitive result is connected with the detail conditions of the graphene sheet. In fact, it has been demonstrated that photoexcitation in a graphene sheet of neutral charge point results in a rise of the carrier concentration and an increase of conductivity, whereas when the initial doping level is high, photoexcitation affects the scattering rate of the carriers that may lead to a conductivity decrease as we observe in our experiment. The feature is connected with the generation of hot carries, the increase of the electronic temperature and the overall increase of the scattering rate observed in highly doped samples. As seen in Figure 1(b), without the photoexcitation the graphene-based structure absorbs a maximum of 75% of the incoming wave at $f$ = 2.17 THz. With photoexcitation the properties of the graphene sheet are modified and the maximum absorption drops; for maximum fluence equal to $I$ = 0.69 mJ/cm$^2$ the absorption drops to 45% providing thus an ultrafast modulation of the absorption of the order of 40%.

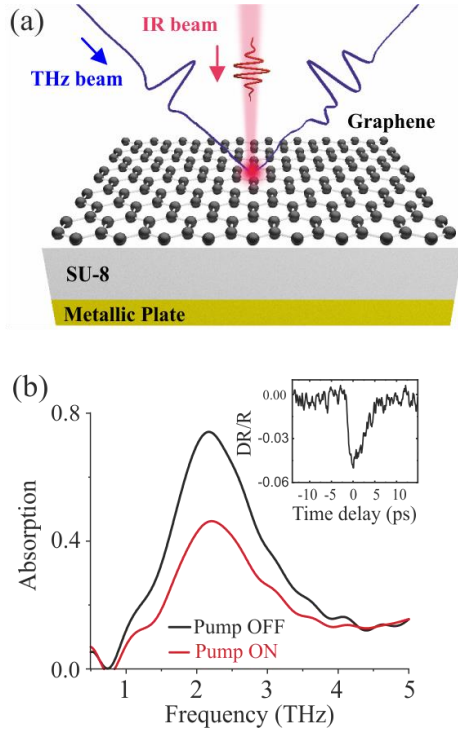

Figure 1: (a) Schematic of the component consisting of a single uniform sheet of graphene placed on top of a back-plated dielectric film and description of modulation operation. (b) THz absorption modulation upon photoexcitation that leads to absorption decrease, inset shows the ultrafast reflectivity modification of the graphene sheet. The decay time is equal to 2.79 ps.

### 3. Conclusions

We studied both experimentally and theoretically a graphene-based thin-film absorber exhibiting ultrafast tunable operation in the THz regime. The structure consists of a uniform graphene sheet grown by CVD on a grounded dielectric substrate; it is designed to provide enhanced absorption based on the coherent interference of the impinging and reflected waves when they are in phase at the lossy graphene sheet at the frequency of 2.17 THz. Our results provide evidence that in our highly doped sample photoexcitation leads to a reduction of the THz conductivity with a decay of 2.79 ps, resulting to an absorption intensity decrease of 40%. We have discussed the dynamics of the photoinduced reduction of the conductivity which is connected with the generation of hot carries, the increase of the electronic temperature and the overall increase of the scattering rate. Our system provides ultrathin, ultrafast modulation appropriate for the demanding future flat optics modulation applications.

### Acknowledgements

This work was supported by the European Union's Horizon 2020 Project 696656 Graphene Flagship, the European Research Council under ERC Advanced Grant no. 320081 (project PHOTOMETA) and the European Union's Horizon 2020 Future Emerging Technologies call (FETOPEN-RIA) under grant agreement no. 736876 (project VISORSURF), the Hellenic Foundation for Research and Innovation (HFRI) and the General Secretariat for Research and Technology (GSRT), under the HFRI PhD Fellowship grant (GA. no. 4894), the National Priorities Research Program grant No. NPRP9 329-1-067 from the Qatar National Research Fund (member of The Qatar Foundation), the H2020 Laserlab-Europe (EC GA 654148) and H2020 MIR-BOSE (EC-GA 737017) projects, Tomsk State University competitive programme and H2020-MSCA RISE-2014 Project ID 644076 CoExAN.